# Large-Scale High PV Power Grid Dynamic Model Development – A Case Study on the U.S. Eastern Interconnection

Shutang You

*Abstract*— Power systems are undergoing a transformation toward a low-carbon non-synchronous generation portfolio. A major concern for system planners and operators is the system dynamics in the high renewable penetration future. Because of the scale of the system and numerous components involved, it is extremely difficult to develop high PV dynamic models based upon actual power system models. The main contribution of this paper is providing an example of developing high PV penetration models based on the validated dynamic model of an actual large-scale power grid — the U.S. Eastern Interconnection system. The displacement of conventional generators by PV is realized by optimization. Combining the PV distribution optimization and the validated dynamic model information, this approach avoids the uncertainties brought about by transmission planning. As the existing dynamic models can be validated by measurements, this approach improves the credibility of the high PV models in representing future power grids. This generic approach can be applied to develop high PV dynamic models for other actual large-scale systems.

## I. Introduction

Power grid reliability is important for almost all sectors of the economy. Many nations have set ambitious targets on renewable energy development, such as wind power and photovoltaic (PV). For example, the U.S. SunShot Initiative goal predicts that solar provide 14% of the U.S. total electricity by 2030 and 27% by 2050 [1, 2]. Assuming that PV generation's load carrying capacity factor is around 0.2, the highest possible instantaneous solar power may reach more than 70% of the total generation [3] for the U.S.. Recent studies on the power grid dynamics under high renewables found many new features of power grids under high penetration of renewables [4, 5]. Furthermore, as the PV penetration is growing, the dynamic issues of power systems, such as transient stability, frequency response, inter-area oscillation, have become major concerns for system operators and planners. Both local controls and wide-area measurement systems have been used to mitigate the impacts of these concerns.

Currently, there are only a few small power grids that have already reached relatively high penetration of renewable in operation. For example, Ireland's operation experience with 50% renewable penetration has indicated the stability issues brought up by wind generation [6]. However, the results and conclusions derived from these small systems may not be directly applicable to large systems such as in U.S. and China.

In addition, every large power grid may have its unique characteristics, such as generation-load distribution, system topology, hydro or thermal dominancy, and load composition. These factors may impact the dynamic features, such as the oscillation damping, and transient stability for different systems under high penetration of renewables. For example, in [7], two different systems: hydro dominant and thermal dominant systems, were used in oscillation impact analysis. It is found that in hydro dominated systems the ultra-capacitor energy storage systems for PV systems is more effective in damping oscillation, while in thermal dominant systems, battery storage systems are more effective. Therefore, even though developing high renewable dynamic models for large-scale power systems is extremely difficult, it is still worthwhile to have a relatively generic approach to develop their high renewable models, based on which mitigation methods can be further developed.

Due to the difficulties of simulating high renewable penetration levels in large-scale power system models, few studies have tried to develop realistic scenarios to simulate high PV levels in actual large-scale systems. The only used method is uniformly or randomly displacing conventional generators by renewables across the system [8]. However, it has been found that the distribution of renewables may have a significant impact on system dynamics [9]. Therefore, it is necessary to project high PV scenarios in a more realistic way for various dynamics studies.

There are many uncertainties in power systems, including but not limited to de-commission of synchronous generators, re-dispatch of power flow, transmission upgrades, and new installation of other renewable plants [10, 11]. Theoretically, all these factors should be properly considered using detailed dynamics planning models that accurately represent a grid in the upcoming five to thirty years. However, given the complexity and economic uncertainty in large-scale power systems, developing accurate future models is out of the question. Even with a series of reasonably developed models, the impact of each factor mentioned above cannot be understood. In addition, in dynamic studies, it is preferable to

This work was primarily supported by the U.S. Department of Energy SunShot National Laboratory Multiyear Partnership (SuNLaMP) under award 30844. This work also made use of the Engineering Research Center Shared Facilities supported by the Engineering Research Center Program of the National Science Foundation and DOE under NSF Award Number EEC-1041877 and the CURENT Industry Partnership Program.

Shutang You is with Department of Electrical Engineering and Computer Science, the University of Tennessee, Knoxville, TN, 37996, USA. (E-mails: syou3@utk.edu).

validate the base-case dynamic model before adding other variations [12]. This is becoming a routine in short-term system dynamic studies [13, 14]. If incorporating those future uncertainties in the dynamic model, the value of model validation would be largely weakened due to additional assumptions and model manipulation.

This study is a pioneer study to project high PV scenarios for developing dynamic models that can represent future large-scale high-PV power systems. This approach combines optimized PV distribution and the validated dynamic model to generate a relatively trustworthy high PV power system dynamic model. More specifically, the dynamic model of the validated case is used as the based case, while the PV distribution is optimized based on expansion economic analysis. The optimization model generates PV distribution considering various system operation and economic factors as well as PV output variations. The approach is demonstrated based on the U.S. Eastern Interconnection system, the largest interconnection in North America. The proposed approach can consider how the economic factors will change the PV distribution and finally system dynamic characteristics. With foreseeing future system dynamic characteristics, the incentive and policy strategies for each market participants could be better developed to avoid additional social costs caused by potential dynamic issues, such as oscillation, inertia inadequacy, voltage and transient stability. It also helps guide the direction of PV grid-integration technology development to enable higher penetration.

The rest of this paper is organized as follows. Section II presents the PV distribution projection and Section III described the high PV dynamic model construction. Section IV demonstrates the implementation of the proposed approach in the U.S. Eastern Interconnection. Section V gives the conclusions.

## II. PV Distribution Projection

The first step of the development of high PV dynamic model is the projection of future PV distribution. This step relates to many factors, including existing generation and transmission infrastructure, load forecasting, solar radiation, fuel price forecast, carbon emission and PV price forecast, and PV sitting land price, etc. In this paper, a modified generation-transmission co-optimization model is proposed to obtain the future PV distribution.

The base generation-transmission co-optimization model was introduced in [10] and realized in PLEXOS. It uses a transportation model to represent the power grid load flow. Since it is mainly used for economic analysis, the economic and operation characteristics of each generator and the interfaces between balance authorities are modeled in details in this study. The generation units inside a region are considered to be connected to a notional node, so the transmission inside a region is not modelled. Nodes that are not associated with a region is considered as regions by themselves. For completeness, the whole model is presented in this section. The modifications on the original model to obtain the PV distribution optimization model will be described in more details.

*1) Objective*

Similar to the routine optimization formulation in PLEXOS [15], the objective of the optimization problem is to minimize the sum of all costs during the planning horizon. The objective consists of three major cost items: the expansion cost, the system operation cost, and the emission cost. The operation cost is the sum of the fixed operation cost, the varying operation cost, the maintenance cost, the penalty of lost load, and the wheeling cost of the transmission network between balance authorities. The expansion cost consists of costs associated with PV expansion, which includes the PV panel price and land costs. The objective function is described in (1).

$$f = C_{\text{PV expansion cost}} + C_{\text{fixed O\&M cost}} + C_{\text{varying O\&M cost}} \\ + C_{\text{fuel cost}} + C_{\text{emission cost}} + C_{\text{wheeling cost}} \\ + C_{\text{cost of lost load}} \quad (1)$$

In this objective function, the PV expansion cost is shown in (2).

$$C_{\text{PV expansion cost}} = \\ \sum_{y=1}^{N_Y} \sum_{r=1}^{N_R} D_y \cdot \left( \left( C_{PVbuilt,r,y} + C_{LandCost,r,y} \right) \cdot x_{PVbuilt,r,y} \right) \quad (2)$$

where $C_{PVbuilt,r,y}$ is the PV panel and installation price. $C_{LandCost,r,y}$ is the land price. $x_{PVbuilt,r,y}$ is PV expansion decision variable. It equals the number of unit of PV installed in region $r$, year $y$. $N_Y$ is the total year number in the study horizon. $N_R$ is the total region number. $D_y$ is the coefficient of cash flow versus present value. Considering "the end year effect", its value over the study horizon is as follows [10, 15]:

$$D_y = \begin{cases} \dfrac{1}{(1+d)^y} & \text{for } y=1,2,\ldots,N_Y-1 \\ \dfrac{1}{(1+d)^{N_Y}} + \dfrac{1}{(1+d)^{N_Y+1}} \cdot \dfrac{1}{\left(1-\dfrac{1}{1+d}\right)} & \text{for } y=N_Y \end{cases} \quad (3)$$

where $d$ is the annual discount rate.

The fixed O&M cost, depending on the rated MW capacity of each unit, is shown in (4).

$$C_{\text{fixed O\&M cost}} = \\ \sum_{y=1}^{N_Y} \sum_{r=1}^{N_R} \sum_{g=1}^{N_{G,r}} D_y \cdot \left[ C_{F\_OM,g,r} \cdot P_{\max,g,r} \left( x_{g,r}^0 + \sum_{y' \leq y} x_{PVbuilt,r,y'} \right) \right] \quad (4)$$

where $C_{F\_OM,g,r}$ is the fixed O&M cost coefficient of unit $g$ located in region $r$. $P_{\max,g,r}$ is the max generation of unit $g$ in region $r$. $N_{G,r}$ is the total unit number (conventional and PV) in region $r$. $x_{g,r}^0$ denotes the existing unit number vector.

To consider the seasonal variation of load, unit generation capability, and renewable resources, each year is divided into several time blocks, each of which is a combination of these variations and forms a steady state of the system operation condition. The length of each block *s* depends on the duration of the operation condition over each year. The operation condition in each block will influence the fuel intake of each generator. Therefore, the cost of fuel in units is shown in (5).

$$C_{\text{fuel cost}} = \sum_{y=1}^{N_Y} \sum_{r=1}^{N_R} D_y \cdot T \sum_{s=1}^{N_{S,y}} DF_{y,s} \cdot \sum_{g=1}^{N_{G,r}} R_{H,g,r} \cdot C_{fuel,g,r,y} \cdot P_{g,r,y,s} \cdot x_{g,r}^0 \quad (5)$$

where $N_{S,y}$ is the total block number in year *y*. $DF_{y,s}$ is the duration fraction of block *s* in year *y*. $R_{H,g,r}$ is the heat rate of unit *g* in region *r*. $C_{fuel,g,r,y}$ is the fuel price for unit *g*, region *r*, year *y*. $P_{g,r,y,s}$ is the output of unit *g* in region *r*, year *y*, block *s*.

The varying O&M cost is shown in (6), representing the incremental cost of each kWh power associated with machine wear and costs of replacement;

$$C_{\text{varying O\&M cost}} = \sum_{y=1}^{N_Y} \sum_{r=1}^{N_R} D_y \cdot T \sum_{s=1}^{N_{S,y}} DF_{y,s} \cdot \sum_{g=1}^{N_{G,r}} C_{V\_OM,g,r} \cdot P_{g,r,y,s} \cdot x_{g,r}^0 \quad (6)$$

where $C_{V\_OM,g,r}$ is the varying O&M cost coefficient of unit *g* in region *r*.

The penalty of unserved load is shown in (7).

$$C_{\text{cost of lost load}} = \sum_{y=1}^{N_Y} \sum_{r=1}^{N_R} D_y \cdot T \sum_{s=1}^{N_{S,y}} DF_{y,s} \cdot C_{VOLL,r} \cdot P_{USP,r,y,s} \quad (7)$$

where $C_{VOLL,r}$ is the penalty price of lost load in region *r*. $P_{USP,r,y,s}$ is the unserved power in region *r*, year *y*, block *s*.

The wheeling cost of interfaces between balancing authorities is shown in (8).

$$C_{\text{wheeling cost}} = \sum_{y=1}^{N_Y} \sum_{l=1}^{N_L} D_y \cdot T \sum_{s=1}^{N_{S,y}} DF_{y,s} \cdot C_{Wheeling,l} \cdot I_{l,y,s} \quad (8)$$

where $C_{Wheeling,l}$ is the wheeling price in interface *l*. $I_{l,y,s}$ is the power flow of transmission interface *l* in block *s*, year *y*.

The cost of emissions from fossil plants is shown in (9).

$$C_{\text{emission cost}} = \sum_{y=1}^{N_Y} \sum_{r=1}^{N_R} \sum_{g=1}^{N_{G,r}} D_y \cdot \left[ C_{emm,g,r,y} \cdot e_{g,r} \cdot T \sum_{s=1}^{N_{S,y}} DF_{y,s} \cdot P_{g,r,y,s} \cdot x_{g,r}^0 \right] \quad (9)$$

where $C_{emm,g,r,y}$ is the emission price for unit *g*, region *r*, year *y*. $e_{g,r}$ is the generation emission coefficient of unit *g* in region *r*.

*2) Constraints*

The optimization model is to minimize the objective function (1), while satisfying constraints shown as follows.

   *a) Regional power balance constraint*

In each time block for each year, the generation, load, and the interface exchange in each region should be balanced. This constraint is expressed as (10), where each item on the left side denotes the generation, unserved load, and the interface exchange, respectively, for all time blocks and all regions.

$$\sum_{g=1}^{N_{r,G}} P_{g,r,y,s} + P_{USP,y,r,s} + \sum_{l \in \Omega_r} I_{l,y,s} = L_{r,y,s} \quad (10)$$

where $L_{r,y,s}$ is the load in region *r*, year *y*, block *s*.

   *b) PV installation speed constraint*

The expansion of PV in each region is constrained due to the limitations in physically resources and the grid integration approval processing. This constraint is shown in (11).

$$x_{PVbuilt,r,y} \leq \overline{x}_{PVbuilt,r,y} \quad (11)$$

where $\overline{x}_{PVbuilt,r,y}$ is the annual PV expansion upper limit in region *r*, year *y*.

   *c) Unit capacity discount due to forced outages and scheduled maintenance*

Due to the forced outages and scheduled maintenance, the available capacity will be smaller than the installed capacity. Therefore, a capacity discount is applied to the installed capacity for reliability and adequacy consideration. This consideration is described in (12), where the maintenance factor $MF_{r,s}$ depends on the load levels as most maintenance activities are scheduled at off-peak seasons.

$$\overline{P}_{\max,g,r,s} \leq \left(1 - F_{MOR,g,r} \cdot MF_{r,s} + F_{FOR,g,r}\right) \cdot P_{\max,g,r} \cdot x_{g,r}^0 \quad (12)$$

where $\overline{P}_{\max,g,r,s}$ is the max generation of unit *g* in region *r*, block *s*, considering maintenance and forced outages.

   *d) Capacity adequacy constraint*

The available generation capacity in each region should be able to provide load, interface flow, and reserve. The amount of reserve should be able to meet the requirement of contingencies of interfaces and units, as described in (13).

$$\overline{P}_{\max,g,r,s} x_{r,g}^0 + P_{g(PV),r,y,s} \left( x_{g(PV),r}^0 + \sum_{y' \leq y} x_{PVbuilt,r,y'} \right) \geq L_{y,r,s} + Rs_{y,r} \quad (13)$$

where $Rs_{r,y}$ is the min reserve margin in region *r* year *y*.

   *e) Interface transmission capacity constraint*

Interface power flow should be within its limit, as shown in (14).

$$-P_{l,\max} \leq I_{l,y,s} \leq P_{l,\max} \quad (14)$$

where $P_{l,\max}$ is the capacity of interface *l*.

   *f) Regional renewable portfolio constraint*

Some regions have firm targets on the renewable portfolio. The renewable generation in these regions are required to be higher than a percentage of the total regional generation, as shown in (15).

$$\sum_{g=1}^{N_{r,G}} P_{\max,g(PV),r}\left(x^0_{g(PV),r} + \sum_{y'\leq y} x_{PVbuilt,r,y'}\right) \geq$$
$$RPS_{r,y} \cdot \sum_{g=1}^{N_{r,G}} P_{\max,g,r} x^0_{g,r} + \quad (15)$$
$$RPS_{r,y} \cdot \sum_{g=1}^{N_{r,G}} P_{\max,g(PV),r}\left(x^0_{g(V),r} + \sum_{y'\leq y} x_{PVbuilt,r,y'}\right)$$

where $P_{\max,g(PV),r}$ is the maximum generation of PV. $x^0_{g(PV),r}$ is the existing PV plant number. $RPS_{r,y}$ is the minimum percentage of PV capacity in region $r$ year $y$.

*g) PV plants output constraint*

The maximum output of solar plants is constrained by the availability of the solar radiation, which are data inputs as time series for each region. The two constraint sets are shown in (16).

$$P_{g(PV),r,y,s} \leq CF_{PV,r,y,s} P_{\max,g(PV),r,y} \quad (16)$$

where $CF_{PV,r,y,s}$ is the capacity factor of a PV plant in region $r$, year $y$, block $s$,

*h) Dynamic model compatibility constraint*

The maximum PV expansion capacity in each region is constrained to be smaller than the regional total generation output in the validated dynamic model. As explained previously, this additional constraint is essential to ensure all PV output scenarios can be realized based on the validated dynamic model in the model construction step.

$$\overline{P}_{\max,g(PV),r}\left(x^0_{g(PV),r} + \sum_{y'\leq y} x_{PVbuilt,r,y'}\right) \leq \sum_{g=1}^{N_{r,G}} \hat{P}_{g,r} \quad (17)$$

where $\hat{P}_{g,r}$ is the output of unit $g$ in region $r$ of the validated dynamic model.

The optimization problem, in which the objective (1) constrained by (10) - (17), is a mixed integer problem and can be solved by commercial MIP solvers.

As previously described, in this optimization formulation, each block $s$ in the optimization model represents a specific steady state of the varying system operation conditions including load, generation resource availability, and solar radiation. As this study focuses on PV distribution, the solar radiation is used as the secondary variable besides load in partition the blocks in each year. This can be called load-solar multi-regional time-block partition in the planning process, which is can be derived from the load-wind block partition described in [11, 16]. As solar radiation has an obvious daily pattern, the load-solar partition will include most blocks with zero PV output during off-peak hours and much larger output in peak hours. This is caused by the correlation between solar radiation and load curves, which is unobvious for the case of wind [17].

## III. HIGH PV DYNAMIC MODEL CONSTRUCTION

After obtaining the generation mix and the PV distribution at the interconnection level in Section II, the PV dynamic model can be incorporated into the validated dynamic model of the power grid. The development of the high PV dynamic model includes four procedures as shown in Figure 1.

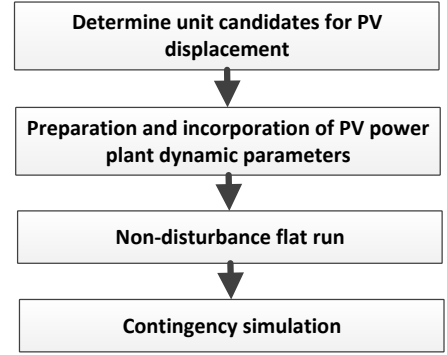

Figure 1. Flowchart of developing high PV dynamic models

*1) Determine unit candidates for PV displacement*

To simulate the high PV scenario for each region, PV displacement candidates are selected according to this sequence: scheduled retiring plants, oil, coal, natural gas, nuclear and hydro. In each region, the PV penetration rate is kept consistent with the regional PV penetration levels determined by Section II.

*2) Preparation and incorporation of PV power plant parameters*

As the high PV power systems are only future-projected scenarios, generic dynamic parameters of PV power plants could be adopted for modelling typical PV plants. In commercial software such as PSS/e, the PV dynamic model can be incorporated into the validated dynamic model through a user-defined model. Proper control modes, such as voltage control and power factor control, should be selected or defined according to simulation purposes before running dynamic simulation test.

*3) Non-disturbance flat run and 4) contingency simulation*

To ensure the developed dynamic models are ready for dynamic studies, dynamic simulation testing on these models is required. In this study, both the twenty seconds non-disturbance scenario and typical N-1 contingency scenarios are tested to validate the numerical stability of the developed dynamic cases. If examining the rotor frequency, a successful flat run will show very small ripples due to very small numerical errors, as shown in Figure 2. A converged contingency run will obtain an expected change in system overall frequency in addition to some common phenomenon, such as electromechanical wave propagation, local and inter-area oscillations, as shown in Figure 3.

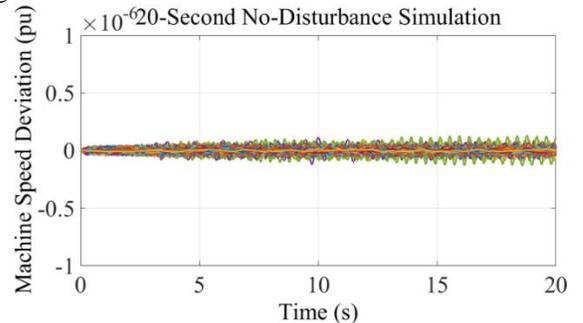

Figure 2. Twenty seconds non-disturbance simulation

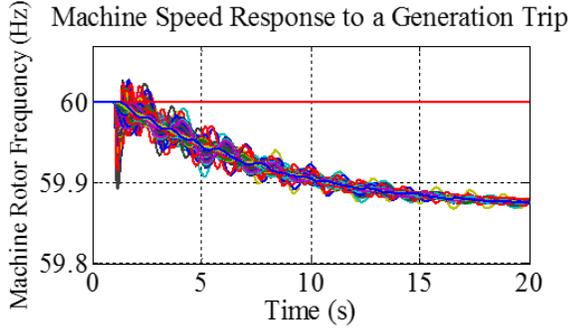

Figure 3. N-1 contingency simulation

## IV. CASE STUDY- U.S. EASTERN INTERCONNECTION

The case study is based on the U.S. Eastern Interconnection (EI) system. The basic statistics information of the EI MMWG (Multiregional Modeling Working Group) model, which is the base-case dynamic model to be used in this study, is given in Table I. As an example, this case study focuses on the impact of PV penetration on system frequency response. For model validation purposes, the base model frequency output and FNET/GridEye measurement at multiple points across the EI were compared using multiple actual events [14, 18]. An example of such comparison at a location in Virginia during a 974 MW generation event is shown in Figure 4. The model sanity check result is given in Table II. It shows that the model is accurate to represent the real frequency response.

To quantify the changes of system frequency response brought about by PV, the PV penetration levels of the to-be-developed simulation scenarios are defined to be 5%, 25%, 45%, and 65%, as shown in Table III.

The proposed PV distribution projection method was applied to optimize the PV distribution in the projected high PV scenarios in the EI system. The horizon featured an interconnection-level PV growth primarily driven by high carbon-emission prices projected in [19]. Input parameters in PV expansion optimization included solar radiation, PV prices, land prices, as well as associated renewable portfolio standards. A summary of input data sources of the PLEXOS model is shown in Table IV.

TABLE I. BASIC INFORMATION OF THE EI MODEL

| EI model Statistics | Value |
|---|---|
| Total Bus Number | 68309 |
| Generator Number | 8337 |
| Branch Number | 58784 |
| Operating Generation Capacity | 600 GW |
| Load | 540 GW |

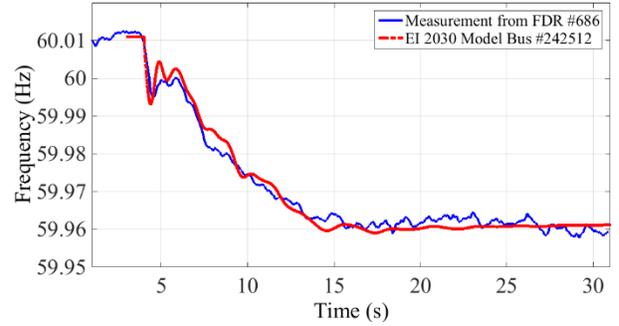

Figure 4. EI model sanity check example

TABLE II. EI MODEL SANITY CHECK METRICS

| | FNET/GridEye Measurement | Simulated Value | Mismatch |
|---|---|---|---|
| Frequency nadir (Hz) | 59.961 | 59.959 | 0.002 |
| Rate of Change of Frequency (mHz/s) | 4.39 | 4.02 | 0.37 |
| Frequency Settling Time (s) | 11.5 | 12.8 | 1.3 |
| Settling Frequency (Hz) | 59.960 | 59.961 | 0.001 |

TABLE III. PV PENETRATION RATES OF ALL SCENARIOS IN EI.

| Scenario | Instantaneous PV Penetration Level |
|---|---|
| Scenario 1 | 5% |
| Scenario 2 | 25% |
| Scenario 3 | 45% |
| Scenario 4 | 65% |

TABLE IV. PLEXOS MODEL INPUT DATA SOURCES

| PLEXOS model input | Data sources |
|---|---|
| • Existing generation and transmission infrastructure<br>• Load forecast<br>• Solar radiation<br>• Fuel price forecast<br>• Carbon emission price forecast | • The Eastern Interconnection Planning Collaborative (EIPC) database [20] |
| • PV price forecast | • North American PV Outlook [21] |
| • PV sitting land price | • Land Value 2015 Summary [22] |

As the main dataset in optimization, the Eastern Interconnection Planning Collaborative (EIPC) Phase I dataset was created by Charles River Associates (CRA) in an effort supported by U.S. Department of Energy involving major stakeholders and planning coordinators in the Eastern Interconnection [19]. It is also compatible with the MMWG dynamic model in representing the current-stage generation-transmission infrastructures. This dataset was translated into

PLEXOS Energy Exemplar for multiple EI expansion studies [23, 24]. It uses a bubble/pipe model representing 24 EI regions (as shown in Figure 5) and the interfaces between them [25]. The colored regions in Figure 5 represent utilities, regional transmission operators, coordinating authorities, independent system operators or other natural groupings based on the structure of the EI.

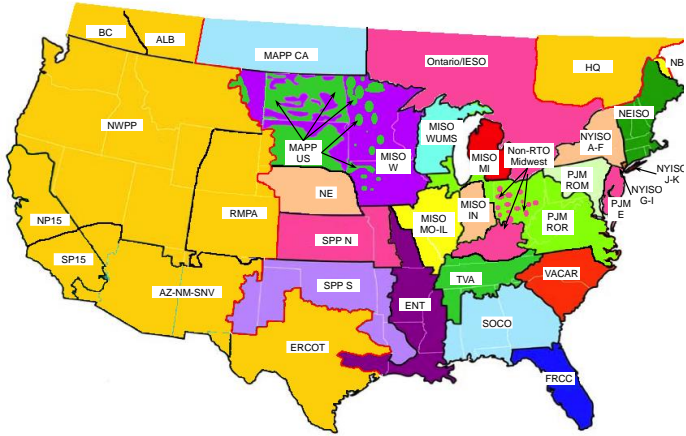

Figure 5. EI Regions in the EIPC Model (excluding all regions in gold)

The commercial mix-integer programming (MIP) solver Xpress-XP was adopted to solve the MIP problem. The PLEXOS optimization result included the generation mix and projected PV capacity distribution for each year in the study horizon, representing multiple PV penetration levels. Based on the PV expansion optimization results, the PV growth in the EI system over the study horizon is shown in Figure 6. It shows a close-to-linearwise increase before some regions' PV has replaced the majority of thermal and nuclear generation. This increase pattern is consistent with some regional PV forecast studies conducted by ISOs and RTOs [26]. Base on the PV growth curve, the PV distribution in four years: 2020, 2024, 2030, and 2035, were selected as the four target penetration scenarios.

An 80% load carrying factor of PV is assumed to simulate a high PV output period in EI. The PV regional instantaneous penetration levels and power outputs for the 25% and 65% PV cases are shown in Figure 7 and Figure 8. The difference between patterns of the penetration level in Figure 7 and power output Figure 8 is due to the difference of the total generation outputs between regions.

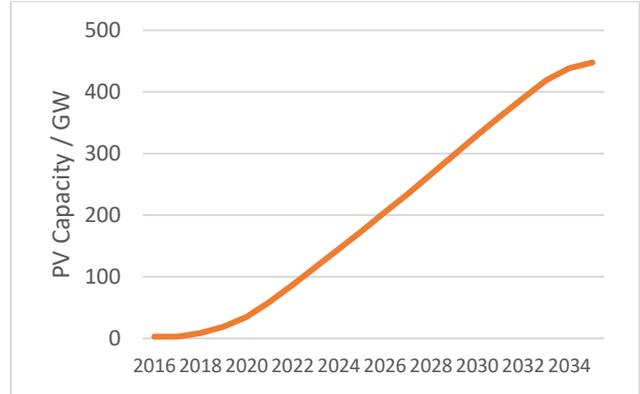

Figure 6. The increase of installed PV capacity in EI during 2015-2035

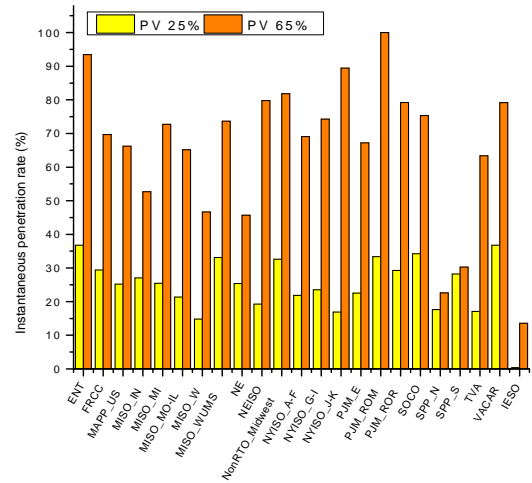

Figure 7. Instantaneous penetration level (25% and 65% PV)

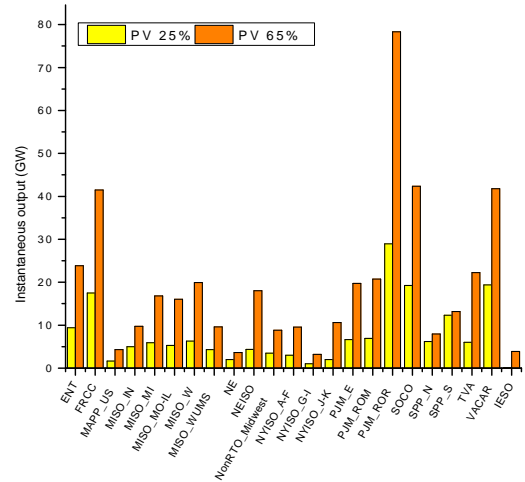

Figure 8. Instantaneous power output (25% and 65% PV)

It can be seen that in the 65% PV case, PJM_ROR has the largest PV output while the PJM_ROM has the highest PV penetration level. Additionally, from a geographic perspective (as shown in Figure 9), the regions in southern EI and those close to load centers tend to have higher PV penetration. That is because the southern regions have higher solar radiation, which leads to higher PV capacity factors and thus more economic benefits by reducing carbon emission and fuel costs, while the regions close to load centers have higher local marginal prices, meaning more economic surplus for the same level of PV output. The location of the PV and other power plants is shown in Figure 10. It can be noted that some regions (such as TVA and MISO) have more conventional power plants than other regions. The reason is that most of these power plants are hydro and wind plants and they are more competitive than thermal plants in terms of fuel and emission costs.

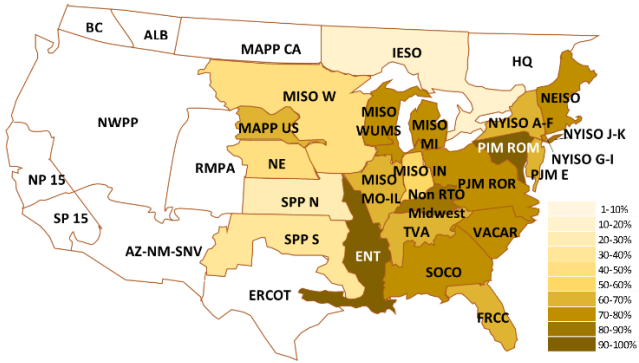

Figure 9. PV penetration distribution (65% PV)

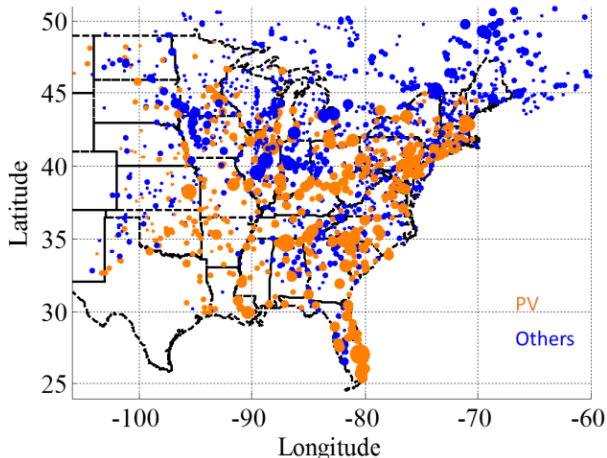

Figure 10. PV and other power plants distribution in the 65% PV case

In dynamic model construction, the high PV dynamic model was constructed based on the validated MMWG dynamic model after incorporating the dynamics of PV power plants according to the projected PV distribution. PV modelling adopted a commonly-used generic PV model as presented in [27]. A 1.2 GW generation trip contingency in TVA was simulated, and the frequency profiles of machine buses in EI are shown in Figure 11. The frequency profiles show the constructed high PV dynamic model has good numerical convergence under this contingency.

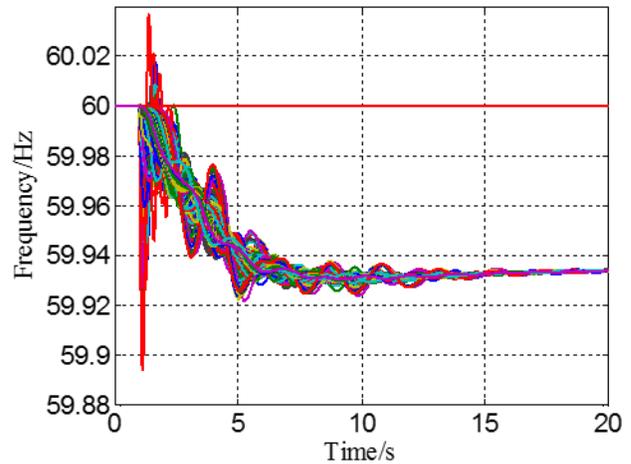

Figure 11. frequency of all machine buses in the 65% PV case

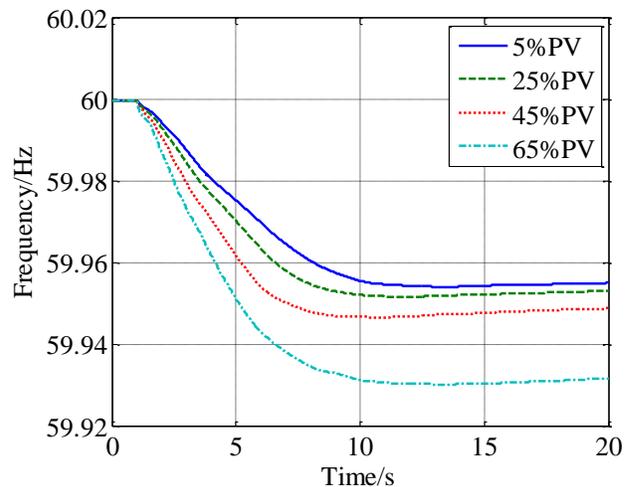

Figure 12. EI frequency response change for different PV penetration rates.

The constructed high PV model can be used for various study purposes for the EI system. Taking frequency response as an example, Figure 12 shows the averaged frequency profiles of the four developed models representing each penetration scenario after the same generation trip disturbance. The 65% PV case shows a significant lower frequency nadir and larger rate of change of frequency due to the reduction of system inertia. This simulation result indicates special attention should be paid to system frequency regulation when the EI system reaches this penetration rate.

## V. CONCLUSIONS

This paper proposed a generic approach to develop high PV system dynamic models for large-scale power system dynamics studies. This approach makes full use the existing validated system dynamic models so that the changes brought up by PV integration can be effectively analyzed and quantified. The PV distribution optimization can consider the various technical-economic factors so that a relatively trustworthy distribution can be obtained. The high PV dynamic model construction procedures are also described in details. The case study on the U.S. EI system and its frequency response under high PV penetration demonstrated the effectiveness of the proposed approach, which is directly applicable to other power grids in studying dynamics under high PV.